# CLMIR: A Textual Dataset for Rumor Identification and Marking

## Authors

Bin Ma, Yifei Zhang, Yongjin Xian, Qi Li, Linna Zhou, Gongxun Miao

## Abstract

With the rise of social media, rumor detection has drawn increasing attention. Although numerous methods have been proposed with the development of rumor classification datasets, they focus on identifying whether a post is a rumor, lacking the ability to mark the specific rumor content. This limitation largely stems from the lack of fine-grained marks in existing datasets. Constructing a rumor dataset with rumor content information marking is of great importance for fine-grained rumor identification. Such a dataset can facilitate practical applications, including rumor tracing, content moderation, and emergency response. Beyond being utilized for overall performance evaluation, this dataset enables the training of rumor detection algorithms to learn content marking, and thus improves their interpretability and reasoning ability, enabling systems to effectively address specific rumor segments. This paper constructs a dataset for rumor detection with fine-grained markings, named CLMIR (Content-Level Marking Dataset for Identifying Rumors). In addition to determining whether a post is a rumor, this dataset further marks the specific content upon which the rumor is based.

## Background & Summary

In the field of rumor detection, several datasets have been developed to facilitate research and model evaluation. The PHEME dataset contains a collection of Twitter rumors and non-rumors posted during breaking news events [1]. It includes data from five events: Charlie Hebdo, Ferguson, Germanwings Crash, Ottawa Shooting, and Sydney Siege. The Twitter15 and Twitter16 datasets focus on tweets from 2015 and 2016, respectively [2][3]. Each dataset consists of source tweets categorized as true rumors, false rumors, unverified rumors, or non-rumors, along with their associated retweets and replies. The LIAR dataset comprises 12,836 short statements from PolitiFact.com, each labeled for veracity across six categories: true, mostly true, half true, mostly false, false, or pants on fire (completely false). This dataset is valuable for studies in fact-checking and fake news detection [4]. For multimodal analysis, the Fakeddit dataset provides over one million Reddit posts, each containing text and image data, categorized into multiple fake news classes [5]. This dataset supports the development of models that process both textual and visual information. Additionally, the LTCR (Long-Text Chinese Rumor) dataset focuses on complex Chinese rumors, particularly those related to COVID-19 [6]. It includes 1,729 real news items and 500 fake news items, each accompanied by detailed metadata, aiding the detection of sophisticated misinformation. COVID19-Health-Rumor [7] focuses on health-related rumors circulating on Chinese social media, particularly Weibo, during the early stages of the COVID-19 pandemic. The dataset includes both the rumors and posts aimed at debunking them, providing insights into how health misinformation spreads and is countered. CHECKED offers a collection of true and false COVID-19 news articles in both JSON and CSV formats, specifically designed for fake news detection research on Chinese social media [8]. CHEF supports evidence-based fact-checking research by providing a Chinese dataset for verifying information with supporting evidence [9]. Combating the Infodemic focuses on identifying and mitigating misinformation related to the COVID-19 outbreak in China, with each record labeled as true, false, or suspicious, providing a valuable resource for infodemic research [10]. The RumourEval[11] dataset supports rumor authenticity detection and stance classification with social media text labeled for credibility and user stance. The Multimodal Lie Detection Dataset combines text and image data for Chinese dialogue-based lie detection[12]. The Weibo Rumor Detection dataset integrates expert-validated rumor samples with text and images for social media rumor detection[13]. FakeNewsNet[14] offers news content, social media context, and spatiotemporal data for studying fake news spread. Snopes Dataset[15], a dataset on real and fake news dissemination analyzes online spread patterns, network characteristics, and user interactions.

Based on these datasets, numerous methods have made notable progress in categorizing rumors. The FakeKG system [16] enhances automated fact-checking by constructing a knowledge graph that represents false claims. This approach significantly improves the speed and accuracy of large-scale fact verification. Another study [17][18][19] focused on the process of multi-hop fact verification, proposing a salience-aware graph learning technique to ensure reliable reasoning. Additionally, the HG-SL model [20] aims at early-stage detection of fake news by jointly learning global and local user propagation behaviors. It combines both wide-ranging user behavior and localized data, thereby improving detection outcomes. The Event-Radar model [21] incorporates multi-view learning into multimodal fake news detection. It introduces an event-driven learning framework, which facilitates the processing and integration of various types of multimodal data. Other research, such as the evidence retrieval method by [22][23][24], highlights the critical role of evidence in verifying facts. They argue that retrieving relevant evidence can almost fully resolve the challenge of fact-checking. Furthermore, evidence-enhanced reasoning frameworks [25][26][27][28] and natural language-based reasoning networks [29][30][31] have significantly contributed to advancing fake news detection, especially in the realm of multimodal detection. Liu et al. [32] explored the transition from skepticism to acceptance by simulating the shifting dynamics of attitudes during the spread of fake news, providing insights into the complex mechanisms driving its dissemination.

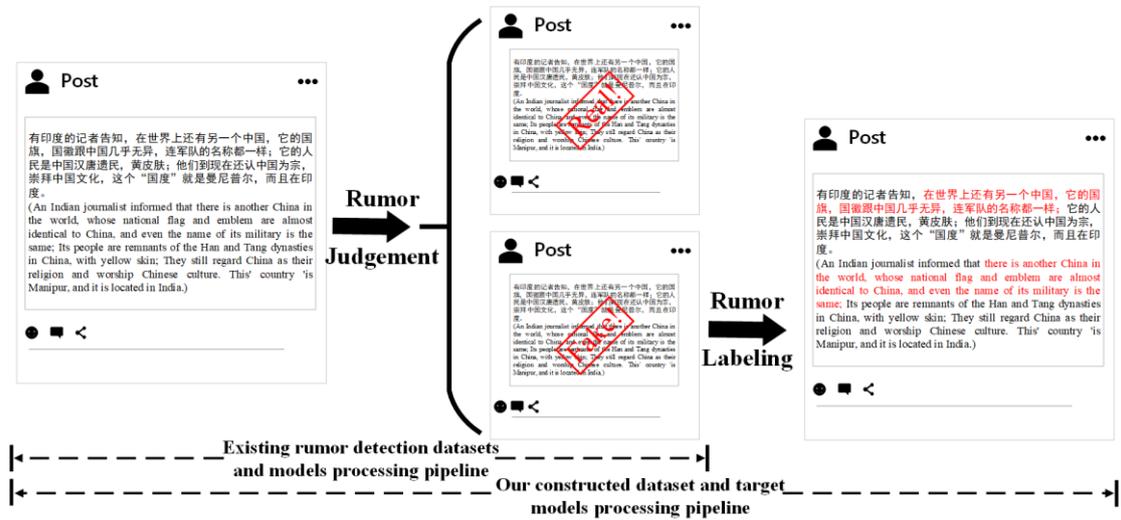

**Figure 1. Comparison between existing and proposed rumor detection datasets**

However, all of these datasets lack the label of specific rumor content. As a result, current rumor detection models primarily focus on classifying contexts as rumors or not, without the capability to mark specific rumor content. The ability to mark specific rumor content is essential for several key reasons. Firstly, it enables a more granular understanding of the rumor itself, facilitating the identification of precise falsehoods or misleading elements within the information. Secondly, this capability allows for the examination of the underlying factors driving rumor propagation, such as the roles of particular entities or the themes that resonate with certain audiences. Additionally, it enhances the interpretability of detection models, providing clearer insights into the components of the content that are identified as false and the rationale behind these determinations.

The importance of a dataset that incorporates labels for specific rumor content becomes increasingly apparent, as it would enable models to move beyond binary classification towards a more nuanced analysis. Such a dataset would not only indicate whether information is a rumor but also highlight the specific elements within the content that are misleading or false. This, in turn, would allow for the development of more advanced models capable of delivering actionable insights and formulating targeted countermeasures.

In this context, we introduce a novel dataset that explicitly addresses this gap, offering labeled examples of specific rumor content. As shown in **Figure 1**, which illustrates a comparison between existing rumor detection datasets and our work, our dataset provides fine-grained markings that pinpoint the exact span of text constituting the rumor. By providing these markings, this work aims to equip rumor detection models with the capacity to not only detect rumors but also to perform detailed analysis of the precise content that constitutes the rumor, thereby advancing the state-of-the-art in rumor detection research.

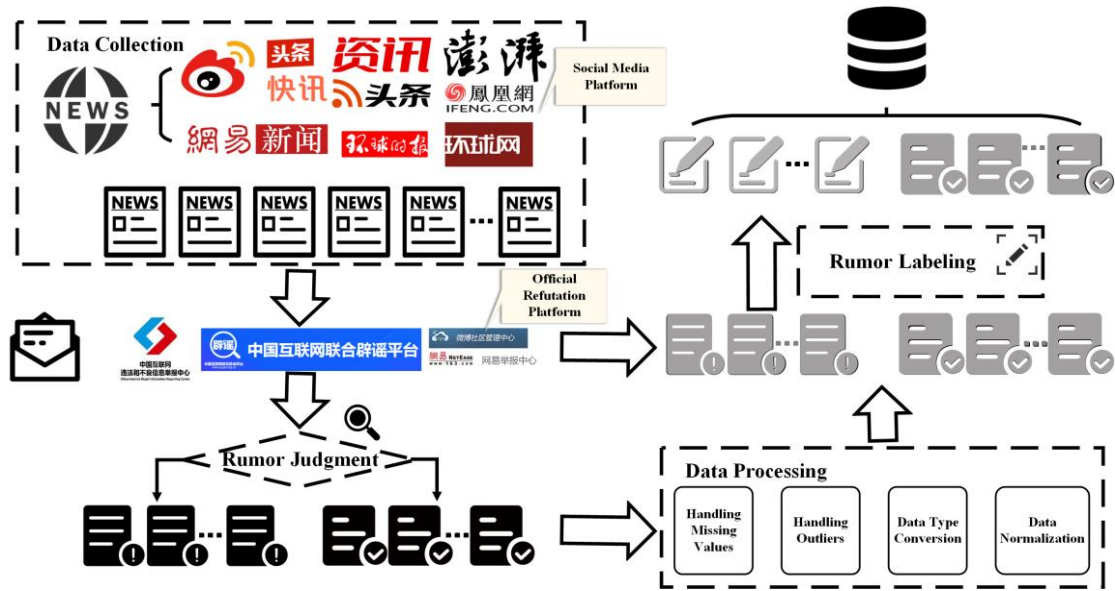

**Figure 2.** Schematic diagram of data processing workflow.

# Methods

We propose a systematic dataset construction process aimed at providing high-quality annotated data for the field of rumor detection. As shown in **Figure 2**, this process consists of four key steps: data collection, data preprocessing, rumor marking, and rumor verification. First, we collect raw data from multiple social media platforms and news websites, such as Weibo, Toutiao, and Tencent News. Next, based on official rumor-refuting platforms and reporting platforms, we make a preliminary judgment on whether the data constitutes a rumor. The judged data is then preprocessed, including handling missing values, outliers, data type conversion, and data standardization, to ensure consistency and usability. Afterward, we design an automated marking algorithm that compares the data with official rumor-refuting platforms (e.g., the China Internet Joint Rumor Refutation Platform) to perform marking. Finally, a manual review is conducted to verify and adjust the marking results, thereby enhancing the accuracy and reliability of the dataset.

## Data Collection.

Rumors exhibit considerable diversity, with distinct characteristics often emerging across different social media platforms. To capture this variability and ensure both diversity and broad coverage in the constructed dataset, we selected a wide range of social media platforms and news websites as data sources. These include Weibo (https://weibo.com), Toutiao (https://www.toutiao.com), The Paper (https://www.thepaper.cn), NetEase News (https://news.163.com), Phoenix News(https://www.ifeng.com), Sohu News (https://news.sohu.com), and Global Times (https://www.huanqiu.com). From these sources, we collected a large volume of news articles and social media posts, laying a comprehensive foundation for robust rumor detection research.

In addition, we collect the confirmed rumor data from authoritative official websites, such as China Internet Joint Rumor Refutation Platform(https://www.piyao.org.cn), Weibo Refutes

Rumors(https://service.account.weibo.com/?type=5&status=4), NetEase Reporting Center(https://jubao.aq.163.com), to ensure the authority of the rumor data source.

## Rumor Judgment.

In order to ensure the authenticity and reliability of the data, after obtaining the data of various social media platforms in the data collection stage, we use the rumor dispelling information released by authoritative rumor dispelling platforms at all levels (such as China Internet Joint Rumor dispelling Platform, People's Daily Online Rumor dispelling Channel, etc.) as the criteria for screening non rumor data, and judge the rumor of the data in this stage. Specifically, we use keyword search to verify relevant information on social media on multiple authoritative debunking platforms, filtering out events or content that have been officially debunked and adding them to the dataset as rumor samples. Events that have not been officially debunked or defined as rumors are added to the dataset as non-rumor samples. The former will form the rumor content of the dataset, together with the rumor data collected from various official debunking platforms.

## Data Processing.

To ensure the integrity and usability of the dataset, a comprehensive data processing pipeline is applied following the rumor judgment phase. It immediately follows "Rumor Judgment" and lays a clean and structured data foundation for "Rumor Labeling" and subsequent model training. This pipeline consists of four key steps: handling missing values, handling outliers, data type conversions, and data normalization.

First, HTML tags, special symbols, and irrelevant formatting characters (such as emojis, hashtags, and excessive punctuation) are stripped from the raw text to reduce redundancy and improve semantic clarity. Next, we remove duplicate entries and normalize textual case by converting all characters to lowercase, thereby eliminating inconsistencies caused by case sensitivity.

In the handling of missing values, samples with missing entries in key fields such as content, timestamp, or source are directly removed to avoid introducing noise or uncertainty into the dataset. Additionally, in the handling of outliers, redundant or irrelevant fields that do not contribute to rumor detection (e.g., formatting tags, advertisement metadata, or unrelated identifiers) are discarded to streamline the data structure and reduce computational overhead.

For data type conversion, all textual and numerical attributes are standardized into consistent and uniform data formats to ensure type consistency across the entire dataset. This includes transforming Boolean fields into binary representations (e.g., 0 and 1). By enforcing strict consistency in data types, we eliminate ambiguities and type mismatches that could otherwise lead to errors during model training or evaluation, thereby enhancing the stability and interoperability of the data processing pipeline.

Data normalization is also applied to ensure consistency across all annotated samples. Specifically, binary labels indicating whether a piece of content is a rumor or not are uniformly mapped to 0 (non-rumor) and 1 (rumor), resolving any discrepancies caused by diverse labeling conventions from different data sources or platforms. This unified format facilitates downstream model training and evaluation by ensuring semantic alignment across the dataset.

# Data Marking.

Figure 3. Rumor Marking Pipeline

To achieve fine-grained recognition of rumor content in social media texts, we have designed a Data Marking process for precise marking of rumor related segments at the span level. This process effectively bridges the gap between the traditional classification task based on whether the entire text is a rumor and the sequence marking task of "where the specific content is", enabling the model to not only determine whether the text is a rumor, but also identify which specific content information is a rumor.

As shown in **Figure 3**, in the previous section, we collected posts from social media platforms that may contain rumors and obtained their official judgments on authoritative debunking platforms. Based on this, we extracted clearly stated rumor statements from debunking information as keywords for matching and marking in the original text. First, we define the labels as follows: B-Rumor indicates the beginning of rumor content, I-Rumor indicates the rumor content, and O indicates non-rumor content. The B-Rumor tag serves to effectively delineate the boundaries of rumors. For instance, when two adjacent sentences in a description both contain rumor content, where the end of the first sentence and the beginning of the second sentence both contain rumor content, they should be labeled as two separate rumor entities, not as a single whole. We designed and applied a rule-based alignment strategy based on this, namely the Hard Matching Tagging Algorithm, as detailed in Algorithm 1. This algorithm searches for the starting position of keywords in the original text and generates a character by character label sequence, where the starting character of each keyword is labeled as B-Rumor and mapped to "0", the middle part is labeled as I-Rumor and mapped to '1', and the remaining non-rumor parts are labeled as O and mapped to '2'. If the keyword does not match successfully in the text, the sample will be discarded to ensure marking accuracy.

This mark mechanism provides a technical foundation for building high-quality segment level rumor datasets, which can be used to train and evaluate sequence marking models such as BiLSTM, CRF, BERT-CRF, or lightweight attention mechanism networks. The clear alignment and marking process further enhances the interpretability and robustness of the rumor detection system in real-world application scenarios.

> **Algorithm 1: Hard Matching Tagging Algorithm**
>
> Input:
>   rumor_text: the original rumor text string
>   keyword: the extracted keyword string
> Output:
>   label_sequence: a sequence of labels corresponding to each character in rumor_text
>               (0 for start of keyword, 1 for inside keyword, 2 for other)
> Procedure:
>   1. If rumor_text or keyword is not a valid string or keyword is empty:
>   2.     Return None
>   3. Let start_idx ← position of keyword in rumor_text using exact match
>   4. If start_idx = -1:
>   5.     Return None
>   6. Initialize label_sequence as a list of '2' with length equal to rumor_text
>   7. Set label_sequence[start_idx] ← '0'    // Mark the start of the keyword
>   8. For i ← 1 to length(keyword) - 1:
>   9.     If start_idx + i < length(rumor_text):
>   10.        Set label_sequence[start_idx + i] ← '1'    // Mark the inside of the keyword
>   11. Return label_sequence as a space-separated string

## Data Records

The dataset consists of a total of 5977 data entries, including 3194 entries that contain rumor content and 2783 entries that do not contain rumor content. The proportions of these two parts are shown in **Figure 5**(a), with rumors accounting for 53.4% and non-rumor posts accounting for 46.6%.

**Figure 4. Example of data fields**

Each data entry includes several fields, which consist of post text, official judgment, whether it contains rumor content, and span-level labels of the rumor content. This dataset is publicly accessible on Figshare, and examples of field descriptors are shown in **Figure 4**. The number of markings for each label is shown in the **Table 1**, labels follow a BIO scheme: B-Rumor (0) for the start of rumor content, I-Rumor (1) for continuation, and O (2) for non-rumor text.

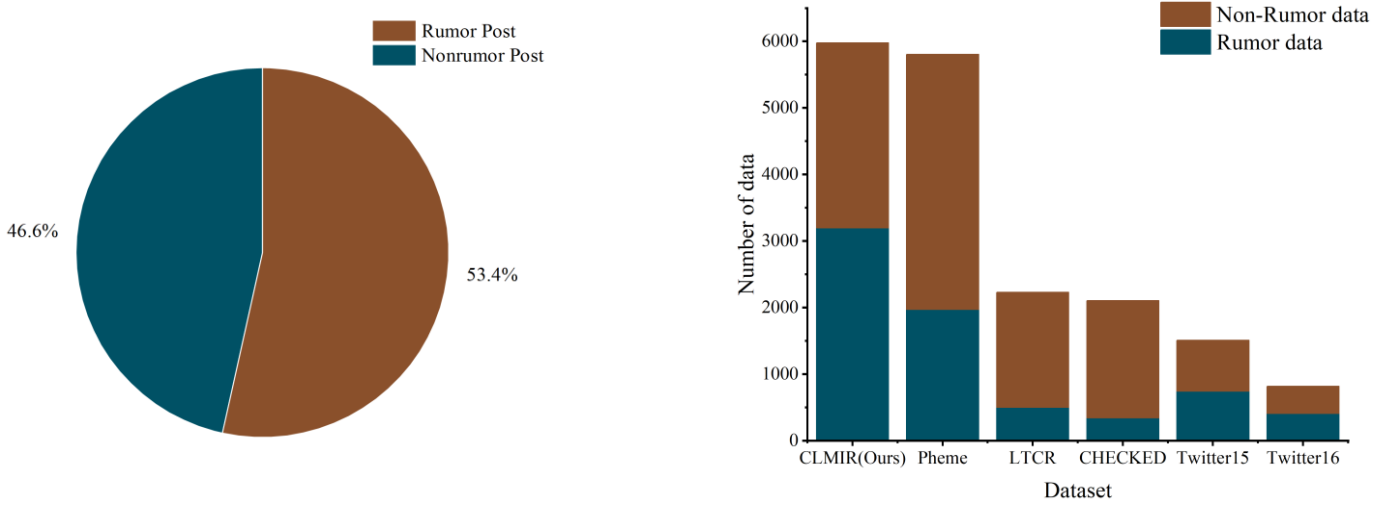

(a) Proportion of rumor and non rumor data in CLMIR        (b) Number of data in the rumor detection dataset

**Figure 5. Analysis of the Quantity of Data in the Dataset**

**Table 1 The statistics of the rumor data in CLMIR. The first row represents the labels, and the second row represents the corresponding number of instances for each label.**

| Statistic | B-Rumors | I-Rumors | O |
|---|---|---|---|
| CLMIR | 3370 | 60409 | 247640 |

## Technical Validation

To ensure the quality and representativeness of our dataset, we performed a comprehensive content analysis, as depicted in **Figure 6**. The analysis highlights several key advantages, affirming the dataset's robustness and its suitability for both rumor detection and broader misinformation research.

CLMIR is built upon a broad and diverse range of data sources, as shown in **Figure 6**(a). It includes content from dynamic social media platforms such as Weibo (35%) and Toutiao (25%), as well as reputable news portals like The Paper, NetEase News, Phoenix News, and Global Times. This diversity allows the dataset to encompass both user-generated discourse and editorially curated content, making it highly representative of real-world online communication. Consequently, it captures the nuanced propagation of rumors across platforms, offering strong generalizability for developing models applicable to varied digital ecosystems.

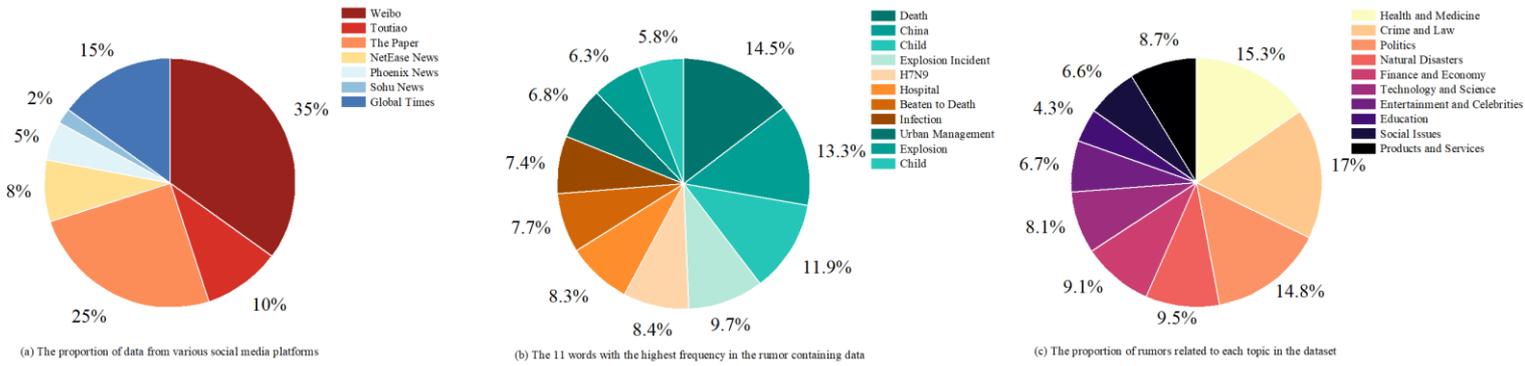

**Figure 6.** Content statistics of the dataset; Among them, Figure (a) shows the proportion of data from various social media platforms, Figure (b) shows the 11 words with the highest frequency in the rumor containing data, and Figure (c) shows the proportion of rumors related to each topic in the dataset.

In addition, the dataset exhibits a semantically rich distribution of keywords, as illustrated in **Figure 6**(b). High-frequency terms such as Death, Child, Explosion, and H7N9 suggest that rumor content often centers around socially sensitive and emotionally resonant topics. This characteristic enriches the dataset with emotional and contextual cues that are beneficial for sentiment-aware misinformation detection. Additionally, the presence of these salient keywords enables interpretable modeling and facilitates the application of attention mechanisms or keyword-driven reasoning frameworks. The dataset thus serves as a valuable resource for studying content virality, emotional triggers, and fear-based information diffusion.

Moreover, **Figure 6**(c) demonstrates that the dataset covers a wide array of topics, including Crime and Law (17%), Health and Medicine (15.3%), and Politics (14.8%), among others. This extensive topical coverage enhances the flexibility of the dataset, supporting multi-label classification and enabling analysis across various social and political domains. It also provides robust training signals for cross-domain generalization and few-shot learning, allowing researchers to develop models that remain effective even in the presence of topic shifts or limited data.

To comprehensively evaluate the effectiveness and representativeness of our proposed CLMIR dataset, we compare it with several widely used rumor detection datasets, including Pheme, LTCR, CHECKED, Twitter15, and Twitter16. As shown in **Figure 5**(b), CLMIR is not only the dataset with the largest number of rumor instances, but also one of the few that exhibit a balanced distribution between rumor and non-rumor content. In contrast, other datasets such as Pheme and CHECKED suffer from significant class imbalance, where non-rumor samples heavily outnumber rumor samples (e.g., Pheme: 1,972 rumors vs. 3,830 non-rumors). This imbalance often results in biased model training and hampers the generalizability of detection models in real-world applications.

In summary, the dataset holds significant technical value and practical application potential. Its well-structured content supports the training and evaluation of advanced deep learning architectures, including transformer-based models, graph neural networks, and multimodal fusion frameworks. Moreover, the event-centric nature of the rumors makes the dataset suitable for tasks such as temporal analysis, causal inference, and event tracking. These properties make it ideal for the development of real-time rumor detection and marking systems, decision-support tools of social media platforms, governmental regulators, and public communication monitoring services.